\documentclass[%
 reprint,
 amsmath,amssymb,
 aps,
]{revtex4-2}
\usepackage{mathptmx}
\usepackage{etoolbox}
\usepackage{xcolor}
\usepackage{float}
\usepackage{hyperref}
\usepackage{graphicx}
\usepackage{dcolumn}
\usepackage{bm}

\begin{document}


\title{Triangular Charge-Density Waves (T-CDW) Stabilize Janus Group-VI Chalcogenide Hydrides}

\author{Jakkapat Seeyangnok$^{1}$}
 \email{jakkapatjtp@gmail.com} 
\author{Udomsilp Pinsook$^{1}$}%
 \email{Udomsilp.P@Chula.ac.th}

\author{Graeme J Ackland$^{2}$}
 \email{gjackland@ed.ac.uk} 
\affiliation{$^{1}$Department of Physics, Faculty of Science, Chulalongkorn University, Bangkok, Thailand.\\
$^{2}$Centre for Science at Extreme Conditions, School of Physics and Astronomy, University of Edinburgh, Edinburgh, United Kingdom}%


\date{\today}

\begin{abstract}
Hydrogenation is an effective strategy for enhancing electron--phonon coupling (EPC) and superconductivity in two-dimensional materials. However, excessively strong EPC can also induce lattice instabilities, leading to charge-density-wave (CDW) formation and structural phase transitions. Here, using first-principles calculations, we investigate CDW order in the Janus transition-metal chalcogenide hydrides 1T-WSH and 1T-WSeH. We find that the high-symmetry phases exhibit pronounced phonon softening at the M point, driving a transition to a commensurate $2\times2$ distorted structure characterized by an emergent triangular charge-density-wave (T-CDW) pattern. Analysis of the electronic structure, susceptibility, and phonon spectrum reveals that the instability is not driven by conventional Fermi-surface nesting but originates from strong momentum-dependent EPC. The T-CDW transition reconstructs the electronic structure and reduces the density of states at the Fermi level, leading to a substantial renormalization of the EPC strength. Consequently, the electron--phonon coupling constants decrease from $\lambda=2.04$ to $1.50$ in 1T-WSH and from $\lambda=3.94$ to $1.06$ in 1T-WSeH, while superconductivity remains robust in CDW phase with predicted transition temperatures of $T_c=12.28$ K and $7.75$ K, respectively. Together with previous results for MoSH and MoSeH, our findings establish a universal mechanism in the 1T-$MCH$ family ($M=\mathrm{Mo},\mathrm{W}$ and $C=\mathrm{S},\mathrm{Se}$), where the primary role of the T-CDW phase is not to eliminate superconductivity but to stabilize the lattice through EPC renormalization. The T-CDW phase therefore acts as an intrinsic self-stabilizing response that relieves excessively strong EPC while preserving phonon-mediated superconductivity.
\end{abstract}

\maketitle

\section*{Introduction}
    Two-dimensional (2D) materials provide a fertile platform for exploring emergent quantum phenomena arising from reduced dimensionality and enhanced many-body interactions. Since the discovery of graphene, atomically thin materials have been found to host a wide variety of collective electronic states, including superconductivity, magnetism, topological phases, and charge density waves (CDWs)~\cite{balandin2021charge}. Among these phenomena, the interplay between superconductivity and CDW order~\cite{wang2023interplay,ali2025interplay} has attracted particular attention because both phases originate from electronic states near the Fermi level and are often intimately connected through electron--phonon coupling (EPC). Understanding how these competing or cooperative orders emerge remains a central challenge in condensed-matter physics.
    
    Hydrogenation has recently emerged as an effective strategy for engineering the electronic and vibrational properties of 2D materials~\cite{bekaert2019hydrogen,seeyangnok2026enhanced,liu2024three,seeyangnok2026stability}. By introducing light hydrogen atoms, the phonon spectrum can be substantially modified and EPC significantly enhanced, leading to phonon-mediated superconductivity with transition temperatures considerably higher than those of pristine materials~\cite{wang2023hydrogenation,seeyangnok2025high,meng2025mbenes,seeyangnok2025ab,seeyangnok2026theoretical,xue2024realization,seeyangnok2025hydrogenation,han2023high,seeyangnok2025phase}. In particular, hydrogenated Janus transition-metal chalcogenides have attracted growing interest following the synthesis of 2H--MoSH~\cite{lu2017janus} and subsequent predictions of superconductivity in MoSH-, MoSeH-, WSH-, and WSeH-based systems~\cite{liu2022two,sui2025two,seeyangnok2024superconductivity,seeyangnok2024superconductivitywseh,qiao2024prediction}. Similar superconducting behavior has also been predicted in several Janus $MX\mathrm{H}$ monolayers ($M=\mathrm{Ti,Zr,Hf}$ and $X=\mathrm{S,Se,Te}$)~\cite{li2024machine,ul2024superconductivity,seeyangnok2025competition}, highlighting a recurring connection between strong EPC and superconductivity in hydrogenated Janus materials.
    
    Strong EPC, however, can also drive lattice instabilities and the emergence of CDW order. A CDW state consists of a periodic modulation of the electronic charge density accompanied by a symmetry-breaking lattice distortion~\cite{gruner2018density}. Although CDWs were originally described within the Peierls framework of Fermi-surface nesting~\cite{peierls1955quantum}, it is now well established that CDW formation in most two-dimensional materials is governed by strongly momentum-dependent EPC and selective phonon softening rather than nesting alone~\cite{johannes2008fermi,zhu2017misconceptions}. This behavior is exemplified by transition-metal dichalcogenides such as NbS$_2$, NbSe$_2$, TaSe$_2$, TaS$_2$, and 2H-Pd$_x$TaSe$_2$, as well as the recently predicted monolayer Mo$_2$NF$_2$, where CDW order remains robust even at the monolayer limit and either coexists or competes with superconductivity~\cite{hwang2024charge,weber2011extended,calandra2011charge,ugeda2016characterization,xi2015strongly,lian2023interplay,bhoi2016interplay,seeyangnok2026competition}.
    
    Recently, EPC-driven lattice instabilities have been reported in  hydrogenated Janus chalcogenides. In particular, first-principles studies of 1T-MoSH and 1T-MoSeH revealed the formation of a commensurate $2\times2$ triangular charge-density-wave (T-CDW) phase~\cite{seeyangnok2026moxhcdw}. Remarkably, the T-CDW distortion stabilizes the lattice by reducing excessive EPC while preserving phonon-mediated superconductivity. These findings raise an important question: is the T-CDW state a material-specific feature of Mo-based compounds, or does it represent a more general self-stabilization mechanism in hydrogenated Janus chalcogenides?
    
    In this work, we investigate the structural, electronic, vibrational, and superconducting properties of the Janus hydrides 1T-WSH and 1T-WSeH using first-principles calculations. Motivated by our recent discovery of an EPC-driven triangular charge-density-wave (T-CDW) phase in 1T-MoSH and 1T-MoSeH~\cite{seeyangnok2026moxhcdw,seeyangnok2026tunable}, we explore whether a similar mechanism operates in the W-based analogues. We show that both compounds exhibit strong phonon instabilities in their high-symmetry phases and spontaneously undergo a transition to a commensurate $2\times2$ triangular CDW state. Through analyses of the electronic structure, susceptibility, phonon spectrum, and electron--phonon coupling, we demonstrate that the instability originates from strong momentum-dependent EPC rather than conventional Fermi-surface nesting. The resulting T-CDW phase lowers the total energy, reconstructs the electronic structure, and significantly renormalizes the EPC strength while preserving robust superconductivity. Our results establish a universal picture for the 1T-$MX$H family ($M=\mathrm{Mo},\mathrm{W}$ and $X=\mathrm{S},\mathrm{Se}$), in which a triangular CDW acts as an intrinsic self-stabilization mechanism in strongly coupled phonon-mediated superconductors.

\section*{Computational Methods}
First-principles calculations were performed within density functional theory (DFT) using the \textsc{Quantum ESPRESSO} package~\cite{giannozzi2009quantum}. Exchange--correlation effects were treated within the generalized gradient approximation (GGA) using the Perdew--Burke--Ernzerhof (PBE) functional~\cite{perdew1996generalized}, together with optimized norm-conserving Vanderbilt (ONCV) pseudopotentials~\cite{schlipf2015optimization}. Plane-wave kinetic-energy and charge-density cutoffs of 80 Ry and 320 Ry were employed, respectively.

The crystal structures were fully optimized until the residual forces on all atoms were below $10^{-5}$~eV/\AA. Brillouin-zone integrations were performed using Monkhorst--Pack $\mathbf{k}$-point meshes~\cite{monkhorst1976special} of $24\times24\times1$ for the primitive cell with Methfessel--Paxton smearing~\cite{methfessel1989high} of 0.02 Ry. For the $2\times2\times1$, $3\times3\times1$, and $4\times4\times1$ supercells, the corresponding $\mathbf{k}$-point meshes were $12\times12\times1$, $9\times9\times1$, and $6\times6\times1$, respectively.

Phonon dispersions, lattice dynamical properties, and electron--phonon coupling (EPC) were evaluated within density functional perturbation theory (DFPT)~\cite{baroni2001phonons}. Dynamical matrices were computed on $12\times12\times1$ and $6\times6\times1$ $\mathbf{q}$-point grids for the primitive and supercell structures, respectively. The interaction between electrons and phonons gives rise to a finite phonon linewidth $\gamma_{\mathbf{q}\nu}$,
\begin{equation} 
\gamma_{\boldsymbol{q}\nu} = 2\pi\omega_{\boldsymbol{q}\nu}\sum_{nm}\sum_{\boldsymbol{k}} \left| g_{\boldsymbol{k}+\boldsymbol{q},\boldsymbol{k}}^{\boldsymbol{q}\nu,mn} \right|^2 \delta(\epsilon_{\boldsymbol{k}+\boldsymbol{q},m}-\epsilon_F) \delta(\epsilon_{\boldsymbol{k},n}-\epsilon_F), \label{gammaphononlinewidths} \end{equation} 

from which the mode-resolved EPC parameter can be expressed as 

\begin{equation} \lambda_{\mathbf{q}\nu}=\frac{\gamma_{\mathbf{q}\nu}}{\pi N(\epsilon_F)\omega_{\mathbf{q}\nu}^{2}}, \end{equation} 

where $N(\epsilon_F)$ is the electronic density of states at the Fermi level and $\omega_{\mathbf{q}\nu}$ denotes the phonon frequency of branch $\nu$ at wave vector $\mathbf{q}$.

The superconducting transition temperature $T_c$ was estimated from the isotropic Eliashberg spectral function $\alpha^2F(\omega)$ using the Allen--Dynes modified McMillan equation~\cite{allen1975transition,pinsook2024analytic},
\begin{equation}
T_c = \frac{f_1 f_2 \omega_{\ln}}{1.2}
\exp\left[
-\frac{1.04(1+\lambda)}
{\lambda-\mu^{\ast}(1+0.62\lambda)}
\right],
\end{equation}
where $\lambda$ is the electron--phonon coupling (EPC) constant, $\mu^{\ast}$ is the Coulomb pseudopotential, and $f_1$ and $f_2$ are the strong-coupling and shape-correction factors, respectively. The EPC constant and logarithmic average phonon frequency were obtained from the Eliashberg spectral function according to
\begin{equation}
\lambda = 2\int_0^{\infty}
\frac{\alpha^2F(\omega)}{\omega},d\omega,
\end{equation}
\begin{equation}
\omega_{\ln}=
\exp!\left[
\frac{2}{\lambda}
\int_0^{\infty}
\frac{\alpha^2F(\omega)}{\omega}
\ln\omega,d\omega
\right].
\end{equation}
Unless otherwise stated, $\mu^{\ast}=0.10$ was adopted throughout this work.


\section*{Emergence of Triangular Charge Density Wave (T-CDW) Order}
    \begin{figure}[h!]
        \centering
        \includegraphics[width=8.5cm]{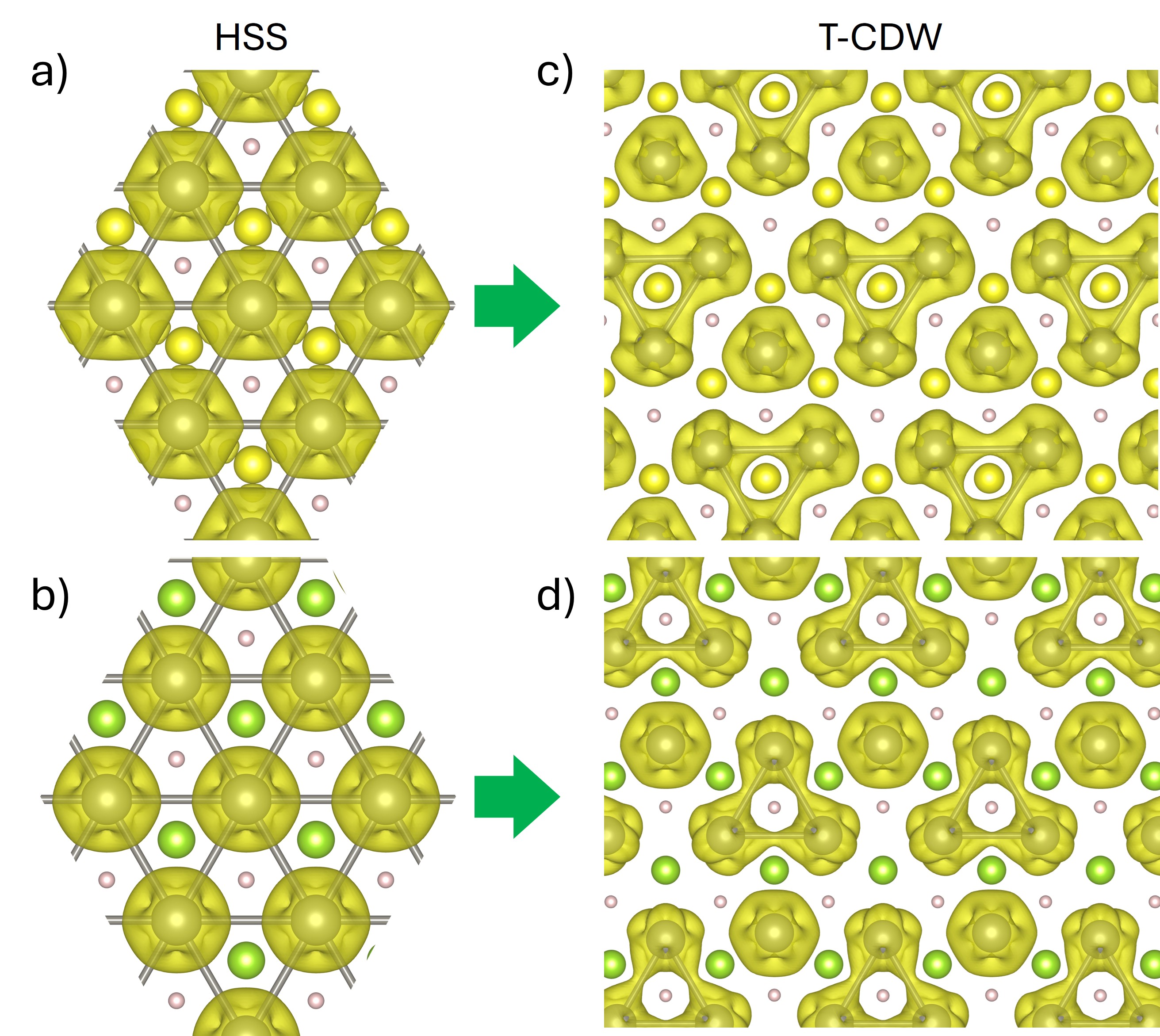}
        \caption{Structural reconstruction and local density ($\rho (r)$) at the Fermi level for 1T-WSH [(a),(c)] and 1T-WSeH [(b),(d)] in the high-symmetry state (HSS) and triangular charge-density-wave (T-CDW) phase, respectively. The T-CDW transition induces a lattice distortion and a strong spatial modulation of the local density $\rho (r)$, giving rise to characteristic triangular electronic textures and a reconstruction of the low-energy states near the Fermi level.}
        \label{fig:LDOS_CDW}
    \end{figure}

    \begin{figure*}[ht]
        \centering
        \includegraphics[width=15cm]{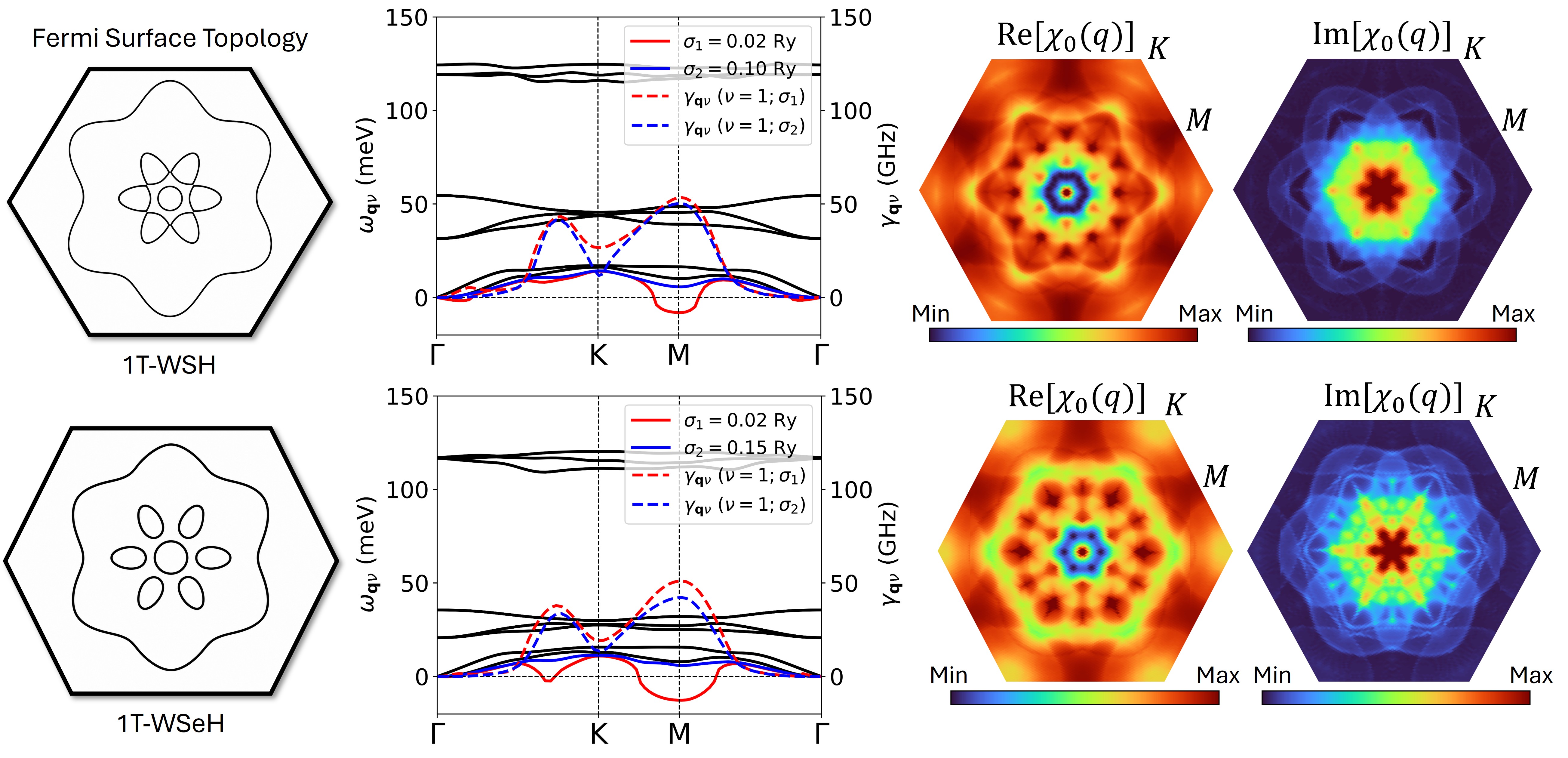}
        \caption{Fermi-surface topology (left), low-energy phonon dispersions (center), and bare electronic susceptibility (right) of the high-symmetry phase of 1T-WSH (top) and 1T-WSeH (bottom). The soft phonon mode near M exhibits a strong dependence on the electronic smearing parameter, demonstrating its electronic origin. Although $\mathrm{Im}[\chi_0(\mathbf{q})]$ shows only moderate enhancement, $\mathrm{Re}[\chi_0(\mathbf{q})]$ develops pronounced maxima near the CDW wave vector, indicating that the instability arises from enhanced electronic susceptibility and electron--phonon coupling rather than perfect Fermi-surface nesting. Together, these signatures reveal the microscopic mechanism driving the transition from the high-symmetry state to the triangular charge-density-wave (T-CDW) phase.}
        \label{fig:HSS_mechanism}
    \end{figure*}

    To characterize the CDW ground state, we first examine the structural and electronic consequences of the transition from the high-symmetry state (HSS) to the distorted phase. As shown in Fig.~\ref{fig:LDOS_CDW}, both 1T-WSH and 1T-WSeH undergo a spontaneous lattice distortion that produces a commensurate $2\times2$ superstructure. The distortion lowers the crystal symmetry and reorganizes the transition-metal atoms into characteristic triangular clusters, giving rise to a triangular charge-density-wave (T-CDW) phase. The transition is driven by the condensation of the soft M-point phonon instability, which removes the dynamical instability present in the HSS and stabilizes the distorted structure. In the CDW phase, three transition-metal atoms undergo sizable in-plane displacements and form a contracted $M_3$ ($M=$ W) triangular cluster, with the W--W bond lengths reduced from approximately 3.07~\AA\ in the HSS to 2.78~\AA\ in the CDW phase. This reconstruction is inconsistent with a simple unidirectional single-$q$ distortion and is instead consistent with a multi-$q$ CDW state arising from the cooperative condensation of symmetry-equivalent M-point phonon instabilities, resulting in the observed triangular CDW pattern. Furthermore, the distorted phase is energetically favored, with energy gains of 9.34 meV/f.u. for 1T-WSH and 37.32 meV/f.u. for 1T-WSeH relative to the HSS, confirming the T-CDW phase as the thermodynamic ground state.

    The structural transition is accompanied by a pronounced reconstruction of the charge density near the Fermi level (local density ($\rho (r)$)). Fig.~\ref{fig:LDOS_CDW}(a,b) show that the local density ($\rho (r)$) in the HSS is distributed relatively uniformly throughout the lattice, reflecting the translational symmetry of the metallic phase. Upon formation of the T-CDW state, however, a strong spatial modulation of the local density ($\rho (r)$) emerges within the enlarged $2\times2$ supercell, mirroring the T-CDW behavior previously identified in 1T-MoSH and 1T-MoSeH~\cite{seeyangnok2026moxhcdw}. As illustrated in Figs.~\ref{fig:LDOS_CDW}(c,d), the low-energy electronic states become concentrated around the triangular structural units generated by the lattice distortion, producing a characteristic triangular electronic texture.

    The close correspondence between the lattice distortion and the local density ($\rho (r)$) modulation demonstrates that the CDW transition involves a cooperative reconstruction of both the atomic and electronic structures. The emergence of triangular structural clusters together with the associated electronic texture provides direct real-space evidence for a T-CDW ground state in 1T-WSH and 1T-WSeH. Similar charge modulations and real-space electronic textures have been reported in layered transition-metal dichalcogenides hosting commensurate CDW order, including NbSe$_2$ and 1T-TaS$_2$~\cite{ugeda2016characterization,weber2011extended,tsen2015structure}. These results highlight the strong coupling between electronic and lattice degrees of freedom mediated by electron--phonon interactions and establish the T-CDW phase as the stable ground state of the Janus transition-metal chalcogenide hydrides.

\section*{Driving Mechanism of T-CDW from the High-Symmetry State}
    To identify the microscopic origin of the T-CDW instability, we examine the interplay between the Fermi-surface topology, electronic susceptibility, and phonon response of the high-symmetry state (HSS). Fig.~\ref{fig:HSS_mechanism} summarizes these quantities for 1T-WSH and 1T-WSeH. Both compounds possess multiple Fermi-surface sheets with sixfold symmetry, suggesting the possibility of electronic instabilities at finite wave vectors. However, the presence of a complex Fermi surface alone is insufficient to establish the driving mechanism of the CDW transition.
    
    To distinguish between a conventional Peierls instability driven by Fermi-surface nesting and an instability originating from electron--phonon interactions, we analyze the bare electronic susceptibility. The real and imaginary parts of the static Lindhard susceptibility are given by
    
    \begin{equation}
    \mathrm{Re},[\chi(\mathbf{q},\omega)] =
    \sum_{\mathbf{k}}
    \frac{f(\varepsilon_{\mathbf{k}})-f(\varepsilon_{\mathbf{k}+\mathbf{q}})}
    {\omega+\varepsilon_{\mathbf{k}}-\varepsilon_{\mathbf{k}+\mathbf{q}}},
    \end{equation}
    
    \begin{equation}
    \mathrm{Im},[\chi(\mathbf{q},\omega)] =
    \pi \sum_{\mathbf{k}}
    \left[f(\varepsilon_{\mathbf{k}})-f(\varepsilon_{\mathbf{k}+\mathbf{q}})\right]
    \delta(\omega+\varepsilon_{\mathbf{k}}-\varepsilon_{\mathbf{k}+\mathbf{q}}),
    \end{equation}
    
    where $\varepsilon_{\mathbf{k}}$ denotes the electronic band energy and $f(\varepsilon)$ is the Fermi--Dirac distribution function.
    
    In the static limit ($\omega\rightarrow0$), Im$[\chi_0(\mathbf{q})]$ reflects the available phase space for Fermi-surface nesting, whereas Re$[\chi_0(\mathbf{q})]$ governs the electronic response that contributes to phonon renormalization. As shown in Fig.~\ref{fig:HSS_mechanism}, Im$[\chi_0(\mathbf{q})]$ exhibits only moderate enhancement near the CDW wave vector and lacks the sharp peak expected for a nesting-driven Peierls transition. In contrast, Re$[\chi_0(\mathbf{q})]$ develops pronounced maxima in the vicinity of the wave vectors associated with the soft phonon modes, indicating a strong momentum dependence of the electronic response.
    
    Additional insight into the origin of the lattice instability is obtained from the electronic-smearing dependence of the phonon dispersions. As shown in Fig.~\ref{fig:HSS_mechanism}, increasing the smearing parameter from $\sigma=0.02$ Ry to $\sigma=0.10$ Ry for 1T-WSH and $\sigma=0.15$ Ry for 1T-WSeH progressively suppresses the phonon softening near the M point and eventually removes the instability. Simultaneously, the phonon linewidths $\gamma_{\mathbf{q}\nu}$ are substantially reduced, indicating a weakening of the electron--phonon interaction. Because electronic smearing broadens the states around the Fermi level and reduces low-energy electron--phonon scattering processes, the pronounced sensitivity of the soft phonon mode to $\sigma$ demonstrates that the instability is closely tied to the low-energy electronic structure. Combined with the absence of a pronounced nesting feature in either Re$[\chi_0(\mathbf{q})]$ or Im$[\chi_0(\mathbf{q})]$, these results rule out a purely nesting-driven mechanism and identify strong momentum-dependent electron--phonon coupling as the primary driving force behind the lattice instability. The enhanced electronic response amplifies the phonon self-energy near the M point, driving phonon softening and ultimately leading to the formation of the commensurate $2\times2$ T-CDW phase.

\section*{Electronic Reconstruction Across the T-CDW Transition}
    \begin{figure}[h!]
        \centering
        \includegraphics[width=8.5cm]{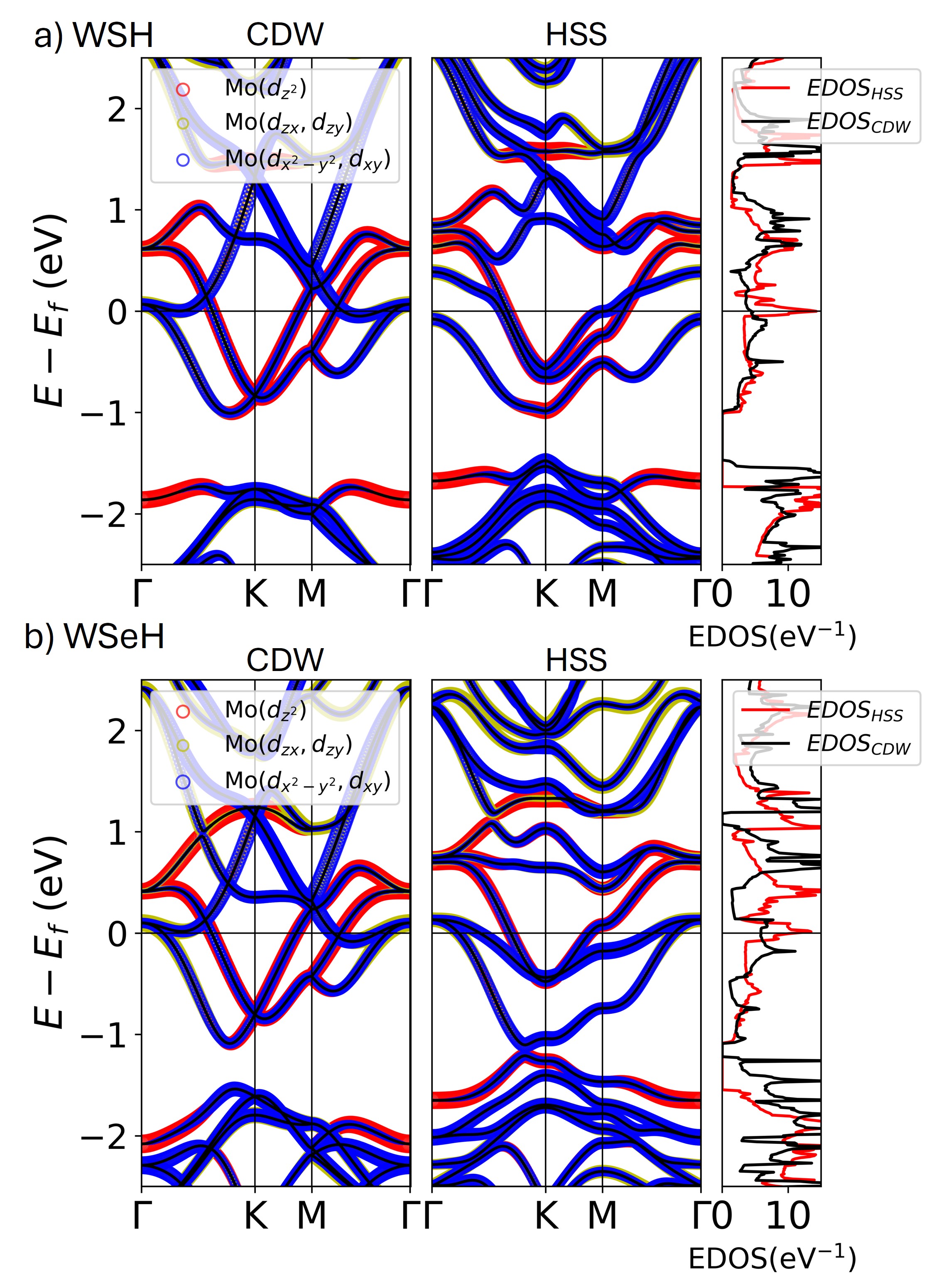}
        \caption{Orbital-resolved electronic band structures in the T-CDW phase (left) and high-symmetry state (center), together with the corresponding electronic density of states (right), for (a) 1T-WSH and (b) 1T-WSeH. The T-CDW transition induces substantial reconstruction of the low-energy electronic structure, including band folding, lifted degeneracies, and a reduced density of states at the Fermi level.}
        \label{fig:EBAND_CDW}
    \end{figure}

     The formation of the T-CDW phase is accompanied by a substantial reconstruction of the electronic structure near the Fermi level. To elucidate the microscopic electronic changes associated with the structural transition, Fig.~\ref{fig:EBAND_CDW} compares the orbital-resolved electronic band structures and electronic density of states (EDOS) of the high-symmetry state (HSS) and T-CDW phase for 1T-WSH and 1T-WSeH. For a direct comparison, both electronic structures were calculated using the same $2\times2\times1$ supercell. In the HSS, the low-energy electronic structure is dominated by transition-metal $d$ orbitals, with significant contributions from the $d_{z^2}$, $(d_{xz},d_{yz})$, and $(d_{x^2-y^2},d_{xy})$ manifolds near the Fermi level. Several bands cross $E_F$, producing a metallic state with a relatively large density of states at the Fermi level.
    
    Upon formation of the T-CDW phase, the electronic bands undergo a pronounced reconstruction. The lattice distortion lifts several band degeneracies and enhances hybridization among low-energy states, particularly near the K and M points. As a consequence, partial gaps develop at multiple band-crossing points, reflecting the strong coupling between the electronic states and the CDW modulation. The reconstructed band structure is accompanied by a pronounced reduction of the electronic density of states near $E_F$ in both compounds. Nevertheless, no full gap opens at the Fermi level, and the systems remain metallic, consistent with a metallic CDW state in which only a subset of Fermi-surface states participates in the reconstruction.
    
    The depletion of low-energy electronic states has important consequences for the electron--phonon interaction. Because the EPC strength depends sensitively on the availability of electronic states near the Fermi level, the reduction of $N(E_F)$ weakens the dominant electron--phonon scattering channels and provides a natural microscopic explanation for the suppression of EPC in the T-CDW phase. Together, the lifted band degeneracies, enhanced orbital hybridization, and reduced density of states at the Fermi level establish a direct connection between the momentum-dependent EPC instability of the HSS and the stabilization of the T-CDW ground state.

\section*{T-CDW-Induced Renormalization of Electron--Phonon Coupling}
    \begin{figure}[h!]
        \centering
        \includegraphics[width=8.5cm]{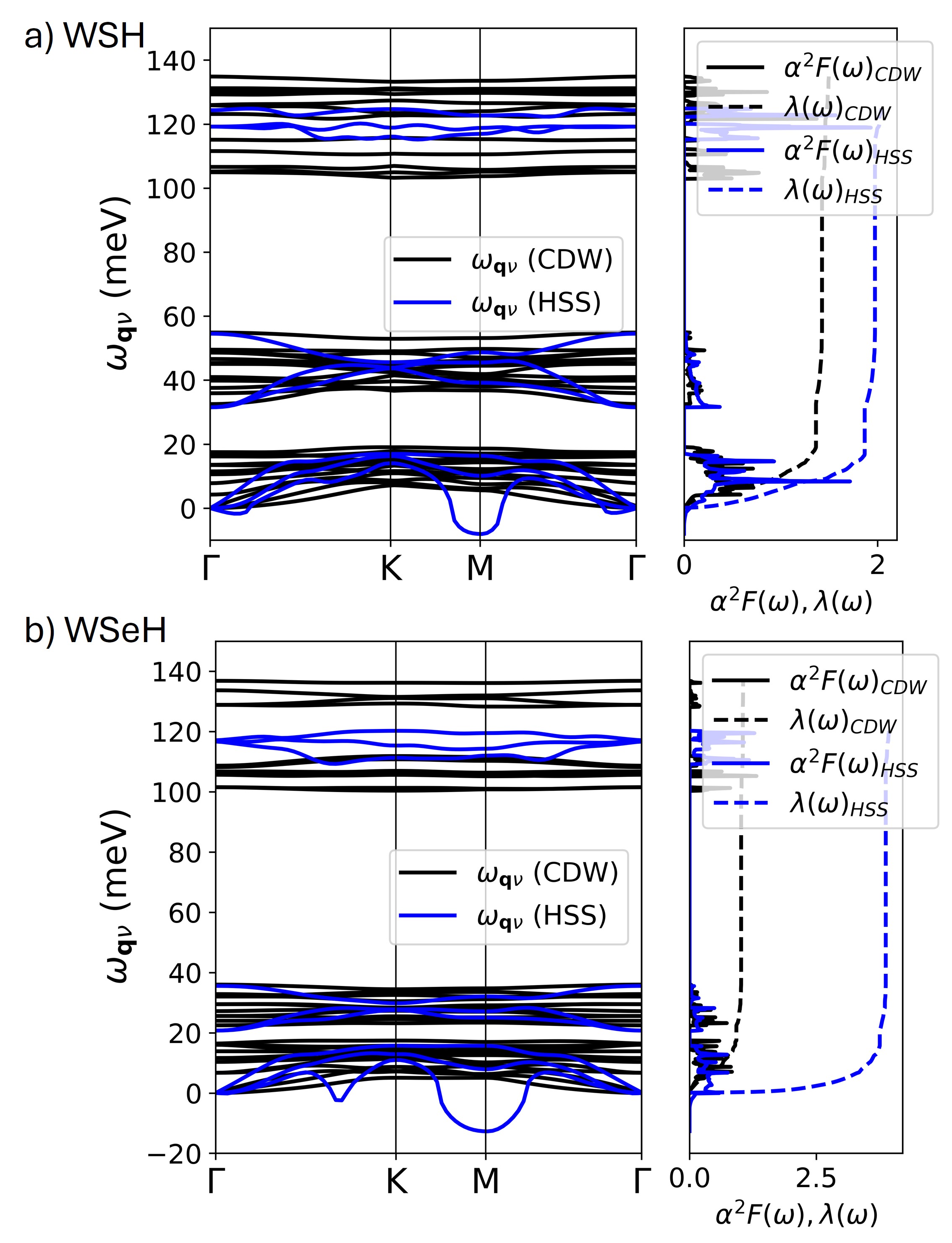}
        \caption{Phonon dispersions (left), Eliashberg spectral functions $\alpha^2F(\omega)$, and cumulative electron--phonon coupling strengths $\lambda(\omega)$ (right) for 1T-WSH (a) and 1T-WSeH (b) in the high-symmetry state (blue) and T-CDW phase (black). The T-CDW transition removes the soft phonon instability and substantially suppresses the electron--phonon coupling, leading to a dynamically stable ground state.}
        \label{fig:EPC_CDW}
    \end{figure}

    The formation of the T-CDW phase profoundly modifies the electron--phonon interaction and provides the microscopic mechanism responsible for stabilizing the lattice. Fig.~\ref{fig:EPC_CDW} compares the phonon dispersions, Eliashberg spectral functions $\alpha^2F(\omega)$, and cumulative electron--phonon coupling strengths $\lambda(\omega)$ of the high-symmetry state (HSS) and T-CDW phase for 1T-WSH and 1T-WSeH. In the HSS, both compounds exhibit pronounced phonon softening near the M point, indicating a dynamical instability driven by strong electron--phonon coupling. The instability is particularly severe in HSS of 1T-WSeH, where the soft phonon branch extends deeply into the imaginary-frequency region. Correspondingly, the EPC constants are exceptionally large, reaching $\lambda=2.04$ and $3.94$ for WSH and WSeH, respectively, indicative of a strong tendency toward structural reconstruction.
    
    The condensation of the soft phonon mode drives the formation of the commensurate $2\times2$ T-CDW phase. In the distorted structure, all phonon frequencies become positive throughout the Brillouin zone, confirming the dynamical stability of the T-CDW ground state. Simultaneously, the electronic reconstruction discussed above reduces the density of states at the Fermi level and suppresses the low-frequency contribution to $\alpha^2F(\omega)$. As a result, the EPC strength is significantly renormalized, decreasing from $\lambda=2.04$ to $1.50$ in 1T-WSH and from $3.94$ to $1.06$ in 1T-WSeH. The corresponding $\lambda(\omega)$ curves clearly show that the T-CDW phase reaches substantially lower saturation values than the HSS, demonstrating that the structural distortion selectively removes the EPC channels associated with the unstable phonon modes.
    
    Despite this substantial EPC reduction, superconductivity remains robust in the T-CDW phase. Using the Allen--Dynes formalism with $\mu^{*}=0.1$, we obtain superconducting transition temperatures of $T_c=12.28$ K and $7.75$ K for WSH and WSeH, respectively. The T-CDW structures are also energetically favored by 9.34 meV/f.u. and 37.32 meV/f.u., confirming that the distorted phases represent the stable ground states. The exceptionally large EPC strengths in the HSS originate from the pronounced softening of low-frequency phonon modes, consistent with the general picture that Kohn-like phonon anomalies can strongly enhance electron--phonon coupling and superconductivity~\cite{jiang2023possible}. In the present systems, however, the softening becomes sufficiently strong to destabilize the lattice and drive the formation of the T-CDW phase. These results demonstrate that the primary role of the T-CDW phase is not to eliminate superconductivity but to stabilize the lattice through EPC renormalization. Together with our previous findings for MoSH and MoSeH, they support a universal mechanism in the 1T-$MCH$ family ($M=\mathrm{Mo},\mathrm{W}$ and $C=\mathrm{S},\mathrm{Se}$), where excessively strong EPC drives a T-CDW instability that subsequently suppresses the instability while preserving phonon-mediated superconductivity.

\section*{Conclusions}
    In summary, we have investigated the origin and role of charge-density-wave order in the Janus transition-metal chalcogenide hydrides 1T-WSH and 1T-WSeH using first-principles calculations. Both compounds exhibit pronounced phonon instabilities in their high-symmetry phases, driving a spontaneous transition to a commensurate $2\times2$ distorted structure characterized by an emergent triangular charge-density-wave (T-CDW) pattern. Through analyses of the Fermi surface, electronic susceptibility, phonon dispersions, and real-space electronic structure, we demonstrate that the instability does not originate from conventional Fermi-surface nesting but is driven by strong momentum-dependent electron--phonon coupling.
    
    The T-CDW transition is accompanied by substantial electronic reconstruction, including lifted band degeneracies, orbital rehybridization, and a reduction of the density of states at the Fermi level. Most importantly, the distorted phase removes the unstable phonon modes and significantly renormalizes the electron--phonon coupling strength, reducing $\lambda$ from 2.04 to 1.50 in 1T-WSH and from 3.94 to 1.06 in 1T-WSeH. Despite this substantial reduction, superconductivity remains robust, with predicted transition temperatures of $T_c=12.28$ K and $7.75$ K, respectively.
    
    Together with our previous studies of MoSH and MoSeH, these results establish a universal mechanism across the 1T-$MCH$ family ($M=\mathrm{Mo},\mathrm{W}$ and $C=\mathrm{S},\mathrm{Se}$), in which excessively strong electron--phonon coupling drives a T-CDW instability that subsequently reconstructs the electronic structure, renormalizes the coupling strength, and stabilizes the ground state. The primary role of the T-CDW phase is therefore not to eliminate superconductivity, but to stabilize the lattice through EPC renormalization. This work provides a unified understanding of the interplay between charge-density-wave order, superconductivity, and electron--phonon coupling in hydrogenated two-dimensional materials and identifies T-CDW formation as an intrinsic self-stabilization mechanism in strongly coupled phonon-mediated superconductors.

\section*{Acknowledgments}
	This work was supported by the Second Century Fund (C2F), Chulalongkorn University (Grant No. C2F PD-2320260067). High-performance computing facility in this Research is funded by Thailand Science research and Innovation Fund Chulalongkorn University (ST-690022300001). This work also made use of the ARCHER2 UK National Supercomputing Service (\url{https://www.archer2.ac.uk}) through the UKCP consortium.

\bibliography{references}

\end{document}